\documentclass{iaus}
\usepackage{graphicx}

\title[HMXB in the SMC] 
{Optical properties of High-Mass X-ray Binaries (HMXB) in the Small Magellanic Cloud }

\author[M.J. Coe et al.]   
{M.J. Coe$^1$, R.H.D. Corbet$^2$, K. E. McGowan$^1$, V. A. McBride$^1$, M.P.E. Schurch$^1$, L.J. Townsend$^1$, J.L. Galache$^3$, I. Negueruela$^4$
 \and D. Buckley$^5$}

\affiliation{$^1$School of Physics \& Astronomy, University of Southampton, SO17 1BJ, UK \\[\affilskip]
$^2$University of Maryland, Baltimore County, Mail Code 662, NASA Goddard Space Flight Center, Greenbelt, MD 20771, USA \\
$^3$Harvard-Smithsonian Center for Astrophysics, 60 Garden Street, Cambridge, MA 02138, USA. \\
$^4$Departamento de Fisica, Ingenieria de Sistemas y Teoria de la Senal, Universidad de Alicante, Apdo. 99, 03080 Alicante, Spain \\
$^5$South African Astronomical Observatory, Observatory, 7935, Cape Town, South Africa}

\pubyear{2008}
\volume{256}  
\pagerange{119--126}
\jname{The Magellanic System: Stars, Gas, and Galaxies}
\editors{Jacco Th. van Loon \& Joana M. Oliveira, eds.}
\begin{document}

\maketitle

\begin{abstract}

The SMC represents an exciting opportunity to observe the direct results of tidal interactions on star birth. One of the best indicators of recent star birth activity is the presence of significant numbers of High-Mass X-ray Binaries (HMXBs) - and the SMC has them in abundance! We present results from nearly 10 years of monitoring these systems plus a wealth of other ground-based optical data. Together they permit us to build a picture of a galaxy with a mass of only a few percent of the Milky Way but with a more extensive HMXB population. However, as often happens, new discoveries lead to some challenging puzzles - where are the other X-ray binaries (eg black hole systems) in the SMC? And why do virtually all the SMC HMXBs have Be star companions? The evidence arising from these extensive optical observations for this apparently unusual stellar evolution are discussed.

\keywords{stars: emission-line, Be; X-rays: binaries}
\end{abstract}

\firstsection 
\section{Introduction}
The Be/X-ray systems represent the largest sub-class of
massive X-ray binaries. A survey of the literature reveals
that of the 115 identified massive X-ray binary pulsar systems
(identified here means exhibiting a coherent X-ray
pulse period), most of the systems fall within this Be counterpart
class of binary. The orbit of the Be star and the
compact object, presumably a neutron star, is generally wide
and eccentric. X-ray outbursts are normally associated with
the passage of the neutron star close to the circumstellar
disk (Okazaki \& Negueruela 2001). A detailed review of the
X-ray properties of such systems may be found in Sasaki et
al. (2003) and a review of the optical properties can
be found in Coe et al. (2005).

\begin{figure}[]
\begin{center}
 \includegraphics[width=2in,angle=-90]{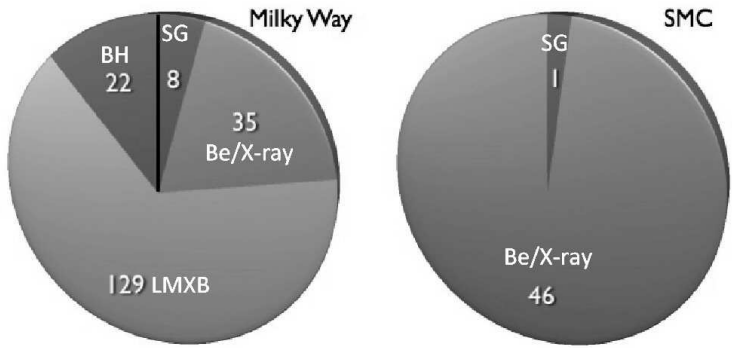}
 \caption{Relative X-ray binary populations in the SMC and the Milky Way.}
   \label{fig1}
\end{center}
\end{figure}

Fig.\,\ref{fig1} shows the current numbers for the different types of X-ray binary populations that are found in the Milky Way and the SMC. Since the number of LMXBs is thought to scale linearly with the mass of hydrogen in the galaxy, then the ratio of $\sim$100 in masses between the two objects explains the lack of LMXBs known in the SMC. But where are the supergiant and Black Hole systems in the SMC?

We currently know of ~40 optically identified systems. This represents by far the largest homogeneous population of X-ray binaries in any galaxy – including the Milky Way. For all these systems OGLE \& MACHO lightcurves exist for $\ge$10 years – enabling the confirmation of counterparts, the identification of binary periods, seeking correlated optical/X-ray flaring etc. These data are supported by follow-up spectroscopy (SALT, AAT \& ESO) –  establishing spectral classes, circumstellar disk status and links into binary evolution

\section{Optical Binary modulation}

\begin {figure}
\begin{minipage}[t]{0.4\linewidth}
\centering
\includegraphics[width=2in]{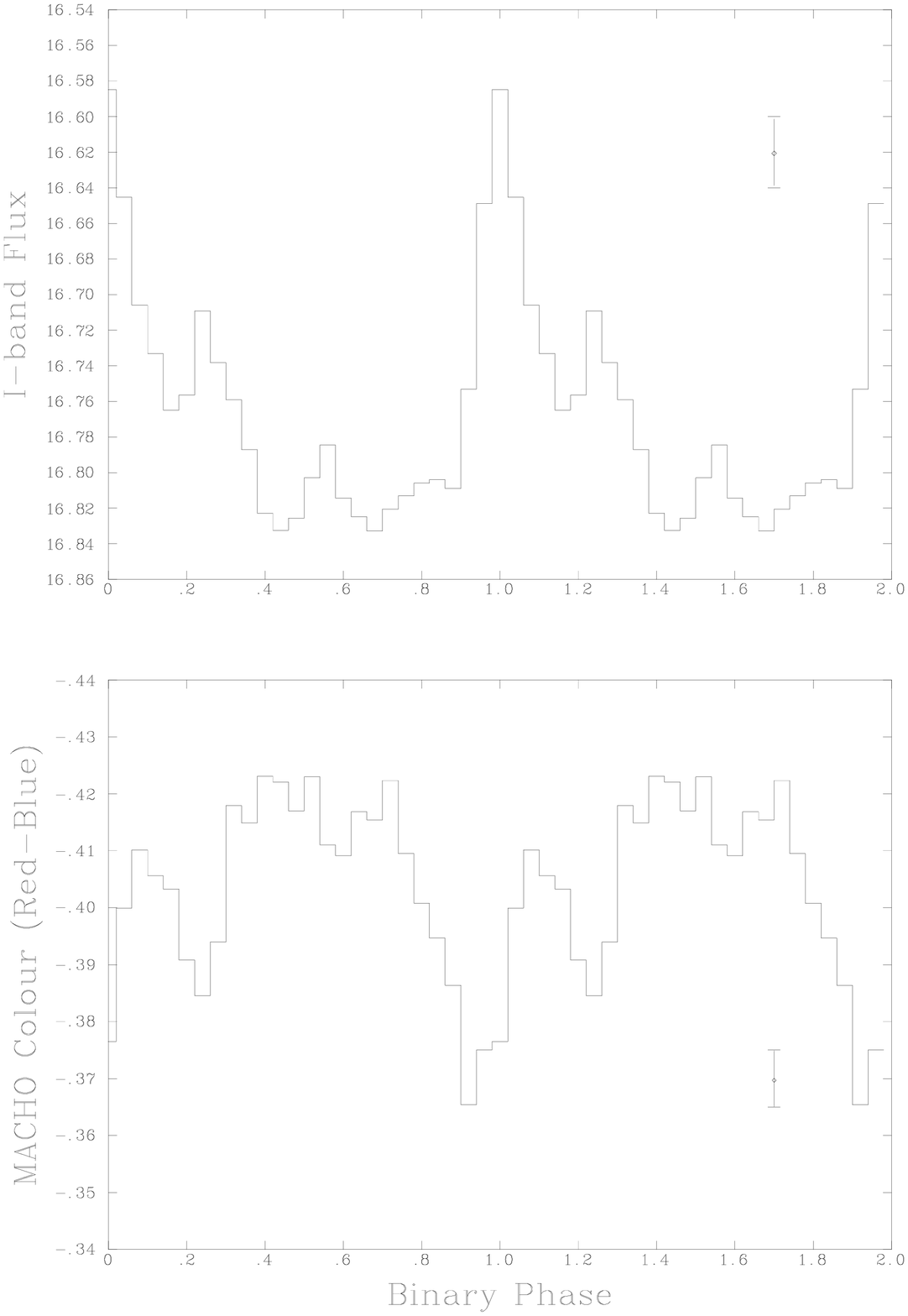}
\caption{Strong correlated colour effects in SXP327 when folded at binary period of 46d.
n.b. the double peaked structure.}
\label{fig2}
\end{minipage}%
\hspace{0.5in}
\begin{minipage}[t]{0.45\linewidth}
\centering
\includegraphics[width=2.5in]{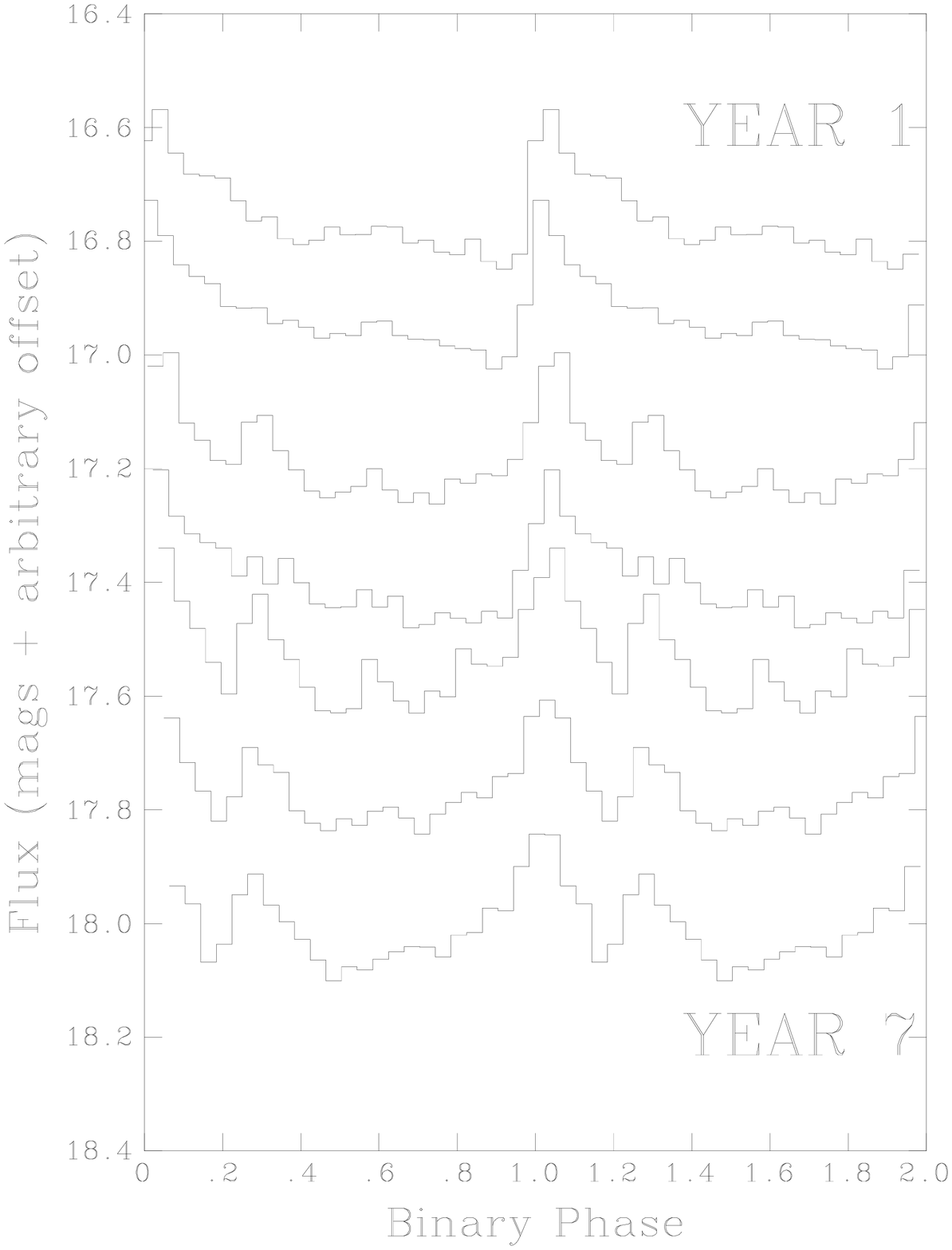}
\caption{Annual variation in the binary profile of the optical
photometry of SXP327 obtained from folding the OGLE III data.}
\label{fig3}
\end{minipage}%
\end{figure}

Some 10-15 of the systems in the SMC have been observed to show a strong optical modulation at the binary period. Some of them also show evidence for quasi-periodic behaviour probably associated with Non Radial Pulsations (see, for example, Schmidtke \& Cowley 2006). SXP327 is an exceptional member of the SMC X-ray binary
pulsar systems in that it shows a very strong optical modulation
at the binary period (Coe et al., 2008) - see Fig.\,\ref{fig2}. Another source, SXP46.6, has also
recently been shown by McGowan et al. (2008) to exhibit optical
flaring at the same phase as X-ray outbursts, but not
in the same strong and consistent manner as SXP327. Those
authors discuss the probable cause of this phenomenon as
lying in the periodic disturbance of the Be stars circumstellar
disk. At the time of periastron passage Okazaki \& Negueruela
(2001) have shown that the disk can be perturbed from
its stable, resonant state with a resulting increase in surface
area and, hence, optical brightness. What is very unusual
about this system, SXP327, is that there is not one, but
at least two outbursts every binary cycle at phases 0.0 and
0.25 (i.e. separated by about 11d). In addition, the average
profile seems to also show a third peak at phase 0.55 - which could be close
to apastron if the main peak represents periastron. Fig.\,\ref{fig2} shows that the colours of the system
reflect the optical brightness. It is obvious from this figure that the
correlation between colour and flux occurs throughout the
binary cycle even though it is most prominent at the time
of the outbursts. The direction of the correlation is to make
the system bluer when brighter - perhaps an indication
of X-ray heating contributing to the colour changes.

On a longer timescale, the average optical modulation varies from year to year (see Fig.\,\ref{fig3}), probably indicative of major changes in the disk structure on the same timescales as the well-known V/R ratio changes.

\section{Population evolution}

There is considerable interest in the evolutionary path of
High-Mass X-ray binary systems (HMXBs), and, in particular,
the proper motion of these systems arising from the
kick velocity imparted when the neutron star was created.
Portegies Zwart (1995) and Van Bever \& Vanbeveren (1997)
investigate the evolutionary paths such systems might take
and invoke kick velocities of the order 100 - 400 km/s. In
an investigation into bow shocks around galactic HMXBs
Huthoff \& Kaper (2002)used Hipparcos proper
motion data used to derive associated space velocities for
Be/X-ray and supergiant systems. From their results, an
average value of 48 km/s is found for the 7 systems that
they were able to fully determine the three dimensional motion.
Since this is rather lower than the theoretical values it
is important to seek other empirical determinations of this
motion.

\begin {figure}
\begin{minipage}[]{0.45\linewidth}
\centering
\includegraphics[width=2.5in]{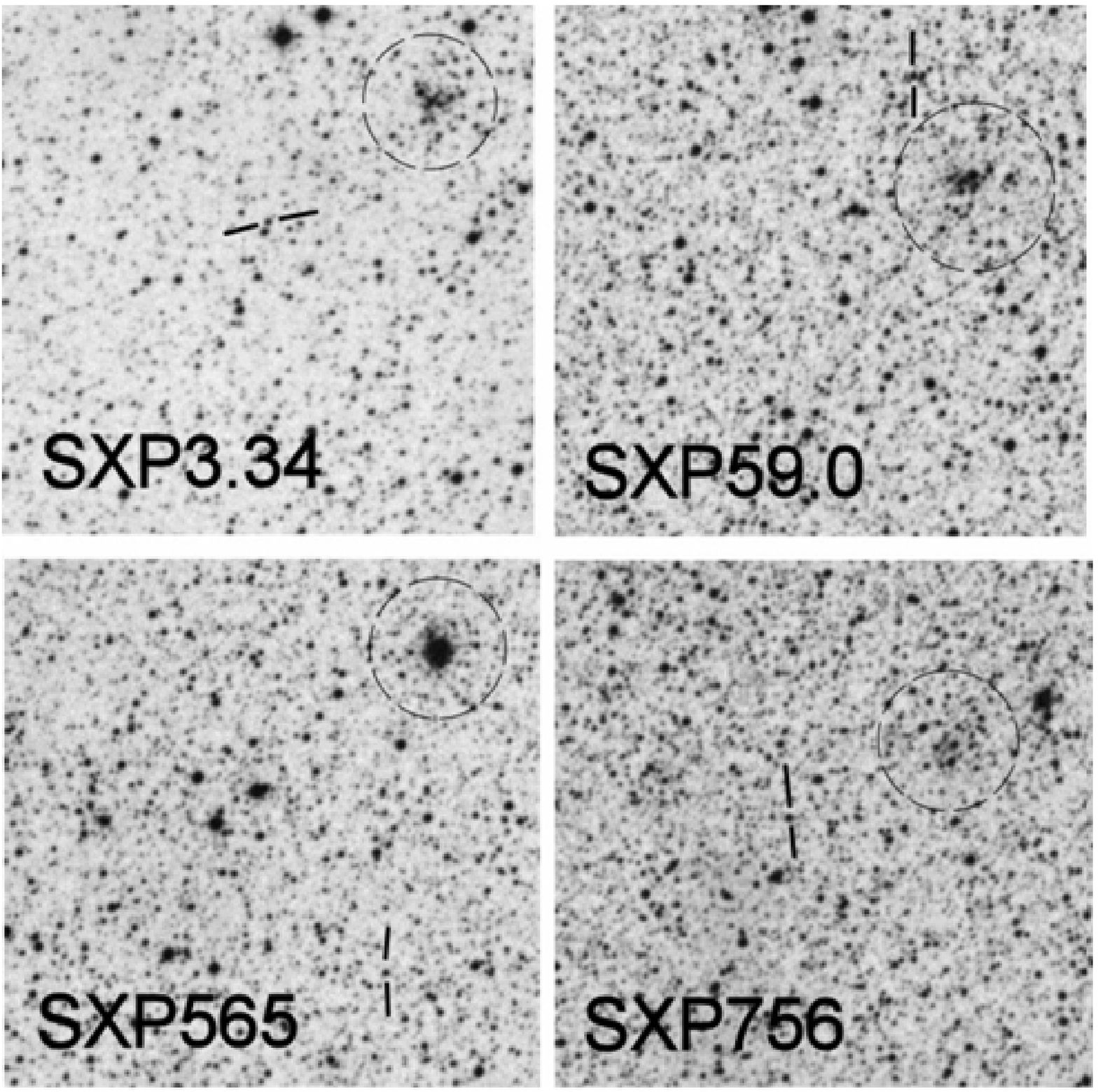}
\caption{The fields around 4 Be/X-ray binary systems as a clue to HMXB evolution. In each case we note the presence of a nearby cluster catalogued by Rafelski \& Zaritsky (2004).}
\label{fig4}
\end{minipage}%
\hspace{0.25in}
\begin{minipage}[]{0.5\linewidth}
\centering
\includegraphics[width=2.2in, angle=-0]{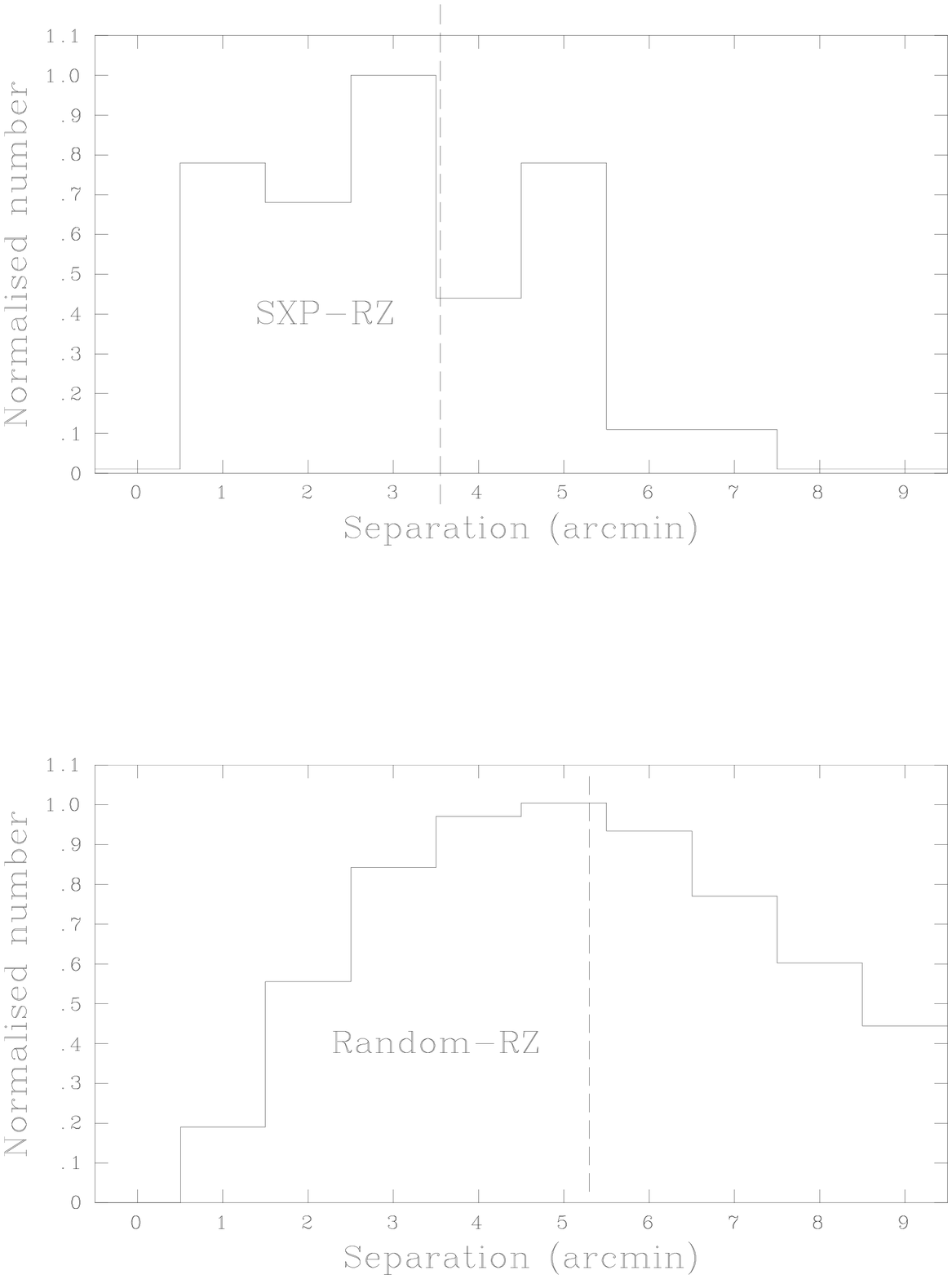}
\caption{Upper panel : histogram of the distances between each
SXP and its nearest neighbour RZ cluster.
Lower panel : histogram
of the minimum distance between 100,000 random points and RZ clusters.
In each case the dotted vertical line shows the position
of the mean of the distribution.}
\label{fig5}
\end{minipage}%
\end{figure}

Coe (2005) used a sample of 17 SMC Be/X-ray binaries to address what may be learnt about kick
velocities by looking at the possible association of HMXBs
in the SMC with the nearby young star clusters from which
they may have emerged as runaway systems (see Fig.\,\ref{fig4}). Here we extend this work to 37 systems.

In order to determine whether the SXP sources may
have originated from a nearby stellar cluster, the coordinates
of the 37 SXP objects were compared to those of the RZ
clusters (Rafelski \& Zaritsky (2005)). For every SXP its position was compared to the location
of all of the RZ clusters and the identification of the
nearest cluster neighbour obtained.
The average distance between the pairs of objects was found to be 3.54 arcminutes. The histogram
of the distances between each SXP source and the nearest
RZ cluster is shown in the upper panel of Fig.\,\ref{fig5}.
Obviously it is important to ensure that the SXP-RZ
cluster distances are significantly closer than a sample of
randomly distributed points. One way to determine this is
simply to just use the RZ cluster data and find the average
cluster-cluster separation. This gives a value of 6.13 arcminutes.
Alternatively, the minimum distance between 100,000
random points and, in each case, the nearest RZ cluster was
found. The average value was found to be 5.30 arcminutes
and the corresponding histogram is shown in the lower panel
of Fig.\,\ref{fig5}. From comparing the two histograms it is clear
that there does exist a much closer connection between SXP
sources and RZ clusters than expected randomly. A paired student t-test of the two distributions gives a probability of only 7\% that the two distributions could have been drawn from the same parent population.

Using a value of 60 kpc for the distance to the SMC, then 3.54 arcminutes corresponds to 60 pc.
Savonije \& van den Heuvel (1977) estimate the maximum possible lifetime of the companion Be star after the creation of the neutron star to be ~5 million years.
So 60 pc indicates the minimum average transverse velocity of the SXP systems is 16 km/s.
van den Heuvel et al. (2000) interpreted the Hipparcos results for galactic HMXBs in terms of models for kick velocities, and obtained values around 15 km/s.

\section{Spectral classification}

With the advent of arcsecond resolution X-ray telescopes
the number of optically identified Be/X-ray binaries
(all but one of the HMXBs in the SMC are Be/X-ray
binaries) in the SMC has risen dramatically over the last
few years. As there are clear differences in the numbers of
HMXBs between the Milky Way and SMC, which can be ascribed
to metallicity and star formation, there may be other
notable differences in the populations. Most fundamentally,
how do the metallicity and star formation rate reflect on
the spectral distribution of the optical counterparts to the
Be/X-ray binary population of the SMC?

Negueruela (1998) showed that the spectral distribution
of Be stars occurring in Be/X-ray binary systems is significantly
different from that of isolated Be stars in the Milky
Way.Whereas isolated Galactic Be stars show a distribution
beginning at the early B-types and continuing through until
A0, the Be star companions of X-ray binaries show a clear
cutoff near spectral type B2.

\begin{figure}[]
\begin{center}
 \includegraphics[width=3.4in]{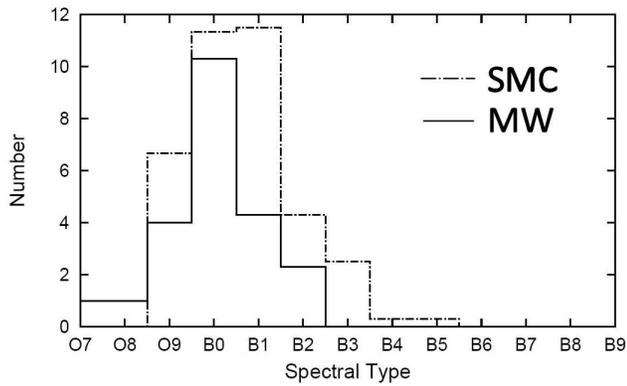}
 \caption{Spectral distribution, determined from blue spectra of $\sim$40 Be/X-ray binaries in the SMC, as compared the distribution of Be/X-ray binaries in the Galaxy. (McBride et al 2008).}
   \label{fig6}
\end{center}
\end{figure}

McBride et al. (2008) carried out detailed blue spectral observations and used these data to classify each counterpart. The spectral distribution of SMC Be/X-ray binaries is shown
in Fig.\,\ref{fig6}. The distribution shows a similarity to the spectral
distributions of the Galactic (Negueruela 1998) and LMC
(Negueruela \& Coe 2002) Be/X-ray binaries. The spectral
distribution of Be/X-ray binary counterparts in the SMC
peaks at spectral type B1, compared to the LMC and Galaxy
distributions, which peak at B0. The Galactic and LMC distributions
show a sharp cutoff at B2, whereas there are 5
SMC objects with possible spectral types beyond B2.
But the exact spectral type cannot
be determined with certainty in these cases. A Kolmogorov-
Smirnov test of the difference between the SMC and Galactic
distributions gives a K-S statistic D = 0.22, indicating
that the null hypothesis (which is that the two distributions
are the same) cannot be rejected even at significances as
low as 90\%. Hence, it is likely that both Galactic and SMC
Be/X-ray binary counterparts are drawn from the same population.

\begin{figure}[]
\begin{center}
 \includegraphics[width=3.4in]{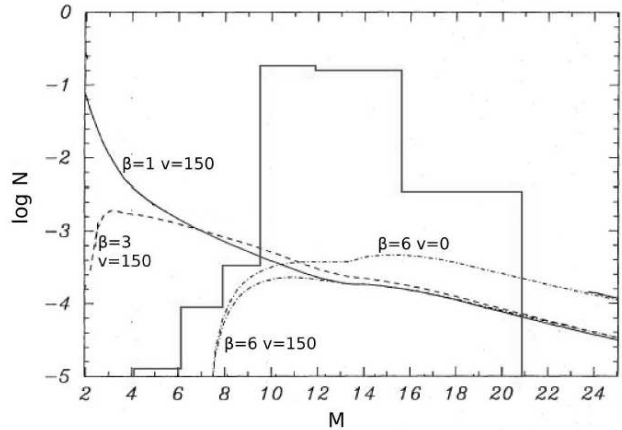}
 \caption{The absolute normalised number distribution of Be
stars with a neutron star companion for four evolutionary scenarios.
The solid histogram represents the spectral distribution of SMC
Be/X-ray binaries from McBride et al (2008).
v represents the supernova kick velocities (in km/s), while $\beta$
represents the amount of angular momentum lost per unit mass loss from the system during evolution.
Original figure from Portegies Zwart (1995).}
   \label{fig7}
\end{center}
\end{figure}

Fig.\,\ref{fig7} shows the arbitrarily scaled spectral distribution
of SMC Be/X-ray binaries superimposed on the
predicted spectral distribution of Be/X-ray binaries (from
Portegies Zwart 1995). As with the Milky Way and LMC
distributions, the SMC distribution cuts off around spectral
type B2 ($\sim$ 8 solar masses), indicating that there may be significant
angular momentum losses in the binary system prior to the
Be/X-ray binary evolutionary phase. A possible interpretation
of the fact that there is no significant metallicity dependence
of the spectral distributions of Be/X-ray binaries
is that the angular momentum is lost through mechanisms
other than the stellar winds of early-type components of
these systems.

\section{Future work \& conclusions}
This next year should provide us with a wealth of new high-energy and optical data of the XRB population in the SMC:

\begin{itemize}
    \item Weekly X-ray (RXTE) monitoring campaign of the SMC Bar will continue for as long as possible.

    \item For the period July - Sep 2008 we will obtain VLT high resolution spectral data on 21 systems every week. We will use the detailed line profiles to study circumstellar disk structures, correlating with X-ray outbursts, as well as RV measurements to identify/confirm binary periods.
    \item From November 2008 till July 2009 ESAs INTEGRAL observatory will carry out a detailed study of the whole of the SMC to a total depth of 2Msec (equivalent to ~70 "nights" of telescope time).
\end{itemize}

So, in summary, we have an excellent homogeneous population of High-Mass X-ray Binaries.
The combined optical \& X-ray data are proving to be a superb laboratory for exploring X-ray binary evolutionary and accretion processes. In addition, the population as a whole has crucial differences with their partners in the Milky Way that need explaining. Finally, the High-Mass X-ray Binaries are providing us with invaluable insights into the recent history of star formation in the SMC.

\end{document}